\documentclass{article}
\usepackage{waspaa23,amsmath,graphicx,url,times}
\usepackage{color}
\usepackage{nameref}
\usepackage{hyperref}
\usepackage{soul}
\usepackage{color}
\usepackage{listings}
\usepackage[symbol]{footmisc}
\usepackage{dsfont}
\usepackage{amsmath}
\usepackage{mathtools}
\usepackage{centernot}
\usepackage{multirow}
\usepackage{makecell}
\usepackage{bm}
\usepackage{amsfonts}
\usepackage{adjustbox}
\usepackage{arydshln}

\usepackage{xpatch}
\let\oldbibitem\bibitem

\newcommand{\newfootnotesize}{\fontsize{8.5pt}{9pt}\selectfont}

\renewcommand{\bibitem}[1]{\oldbibitem{#1}\newfootnotesize}

\usepackage{url}
\newcommand{\footurl}[1]{\begingroup\urlstyle{same}\url{#1}\endgroup}

\DeclarePairedDelimiter\abs{\lvert}{\rvert}%
\makeatletter
\let\oldabs\abs
\def\abs{\@ifstar{\oldabs}{\oldabs*}}

\newcommand{\norm}[1]{\left\lVert#1\right\rVert}

\newcommand{\parn}[1]{\left(#1\right)}

\newlength{\NOTskip}

\title{SLMGAN: Exploiting Speech Language Model Representations for Unsupervised Zero-Shot Voice Conversion in GANs}

\name{Yinghao Aaron Li, Cong Han, Nima Mesgarani}
\address{Department of Electrical Engineering, Columbia University, USA}

\begin{document}

\ninept
\maketitle

\begin{sloppy}

\begin{abstract}
  In recent years, large-scale pre-trained speech language models (SLMs) have demonstrated remarkable advancements in various generative speech modeling applications, such as text-to-speech synthesis, voice conversion, and speech enhancement. These applications typically involve mapping text or speech inputs to pre-trained SLM representations, from which target speech is decoded. This paper introduces a new approach, SLMGAN, to leverage SLM representations for discriminative tasks within the generative adversarial network (GAN) framework, specifically for voice conversion. Building upon StarGANv2-VC, we add our novel SLM-based WavLM discriminators on top of the mel-based discriminators along with our newly designed SLM feature matching loss function, resulting in an unsupervised zero-shot voice conversion system that does not require text labels during training. Subjective evaluation results show that SLMGAN outperforms existing state-of-the-art zero-shot voice conversion models in terms of naturalness and achieves comparable similarity,  highlighting the potential of SLM-based discriminators for related applications.
\end{abstract}

\begin{keywords}
Voice conversion, large language model, generative adversarial networks
\end{keywords}

\section{Introduction}
\label{sec:intro}

Voice conversion (VC), a technique of converting one speaker's voice to another speaker's voice, has gained increasing attention in recent years due to its numerous applications, such as personalized text-to-speech synthesis, speaker anonymization, and entertainment \cite{sisman2020overview}. One type of VC, zero-shot voice conversion, also known as any-to-any voice conversion, has become particularly popular. It aims to convert a source speaker's voice to a target without paired training data and without restricting source and target speakers already seen during training \cite{walczyna2023overview}. There are mainly two approaches to achieving zero-shot voice conversion: reconstruction-based methods and GAN-based methods. 

Reconstruction-based methods focus on disentangling speaker information from linguistic information in the latent representation of speech. There are several approaches in this category, including autoencoder-based approaches \cite{qian2019autovc, yuan2021improving, wang2021vqmivc, chen2021again, lian2022robust}, where input speech is mapped to a bottleneck representation with much smaller dimensions than the input and decoded back to that of the target speaker; ASR-based and TTS-based approaches \cite{casanova2022yourtts, levkovitch2022zero, li2023styletts, hussain2023ace}, where input speech is mapped to the same latent representation as the phoneme representation from the text; and large pre-trained speech language model (SLM) approaches \cite{choi2021neural, qian2022contentvec, dang2022training}, where speech is directly reconstructed from SLM representations in deep layers that contain less speaker information. SLM approaches are particularly popular due to their supreme performance from large pre-training. While these approaches can be effective, they do not guarantee perfect disentanglement of speaker and linguistic information, resulting in unnatural speech or residual source speaker information in the converted speech. Moreover, SLM approaches are often slow in inference because of  the sheer amount of parameters in the large SLMs, making them unappealing for real-time applications.  

GAN-based methods \cite{zhang2020gazev, nguyen2022nvc, takahashi2022robust}, on the other hand, employ a discriminator to determine whether the converted voice is from the target speaker or not. This approach does not necessarily disentangle speaker information from linguistic information, as it relies on the discriminator's ability to capture the human perception of speaker identity. GAN-based methods typically produce more natural speech \cite{levkovitch2022zero, yasur2023deepfake}, as the latent representations are not forced to be disentangled, preserving more linguistic information. However, the similarity of converted speech relies heavily on the discriminative power of the discriminators, yet designing a discriminator that effectively captures each target speaker's speaking style is challenging, especially when the number of speakers in the dataset is large. 

Our recent work has demonstrated promising results by employing SLMs as discriminators for text-to-speech (TTS), where we show that leveraging SLMs as discriminators improves the naturalness of synthesized speech, specifically from paralinguistic aspects \cite{li2023styletts2}. Still, relevant applications in cross-domain transfer, such as voice conversion, have not been examined. In this work, we introduce a novel method to improve the performance of discriminators using speech language models in GAN-based zero-shot voice conversion models involving a large number of speakers. Our approach extends the state-of-the-art StarGANv2-VC \cite{li2021starganv2} model by incorporating a set of SLM-based discriminators in addition to the existing mel-based discriminators. We employ a neural vocoder BIGVGAN \cite{lee2022bigvgan} to convert the generated mel-spectrograms into waveforms, which are then fed into the WavLM \cite{chen2022wavlm} encoder to obtain the SLM representations of the converted speech. Following this, we train discriminators that consist of simple convolutional neural networks (CNN) on these SLM representations to differentiate between the real and fake samples and classify the source speakers. 

Our model retains the advantages of the original StarGANv2-VC model, such as eliminating the need for text labels, enabling potential real-time applications, and exhibiting greater adaptability to other voice conversion tasks like cross-lingual conversion, while benefiting from rich representations learned by self-supervised training in large SLMs like SLM-based approaches do. Additionally, unlike the vanilla StarGANv2-VC, our model scales to a larger number of speakers without any performance loss. Subjective evaluations show that our model surpasses two baseline models, VQMIVIC \cite{wang2021vqmivc} and AGAIN-VC \cite{chen2021again}, in terms of both naturalness and similarity. Furthermore, it outperforms YourTTS \cite{casanova2022yourtts} and StyleTTS-VC \cite{li2023styletts} in terms of naturalness, two of the best-performing publicly available models for zero-shot voice conversion that rely on text labels, albeit no text labels are required during training. The audio samples can be found at \url{https://slmgan.github.io/}.

\begin{figure*}[t]
  \centering
  \includegraphics[width=\linewidth]{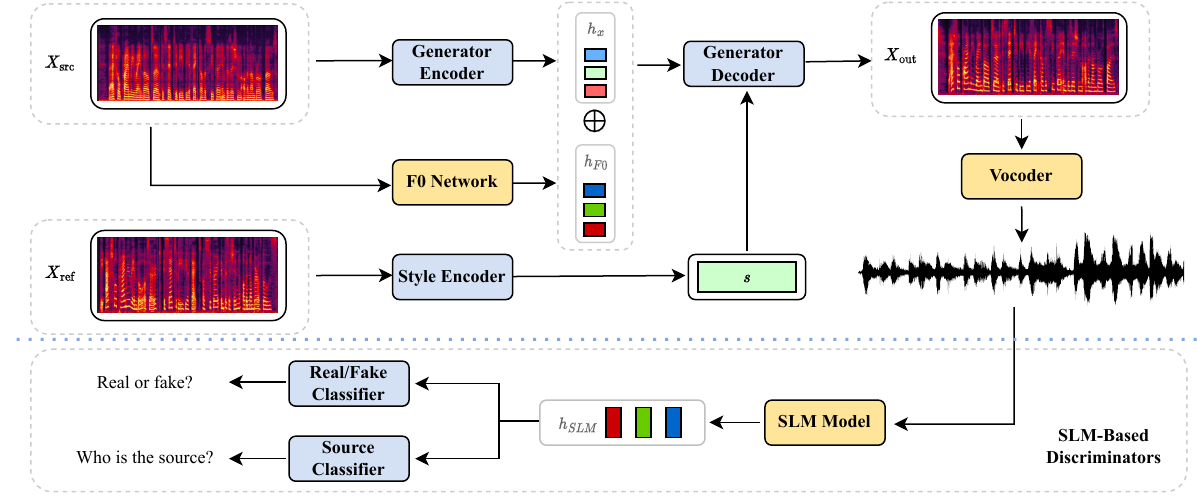}
  \caption{Overview of the training and inference scheme of SLMGAN. Modules in orange are pre-trained and remain fixed during training, while modules in blue are trained. \textbf{Top:} Inference scheme of SLMGAN. $\bm{X}_{\text{src}}$ represents the source input, $\bm{X}_{\text{ref}}$ denotes the reference speech of the target speaker, and $\bm{X}_{\text{out}}$ is the converted mel-spectrogram. $h_x$, $h_{F0}$, and $s$ correspond to the latent feature of the source encoded by the generator's encoder, the F0 feature extracted by the F0 network, and the style vector of the reference speech, respectively. $h_x$ and $h_{F0}$ are concatenated by channel as input to the decoder, while $s$ is injected into the decoder via adaptive instance normalization (AdaIN) \cite{huang2017arbitrary}. The output $\bm{X}_{\text{out}}$ is then converted by a neural vocoder into the waveform $V(\bm{X}_{\text{out}})$. \textbf{Bottom:} Training procedure of SLMGAN with SLM-based discriminators. The pre-trained SLM model extracts the SLM representations $h_{SLM}$ from the waveform, which are subsequently fed into two classifiers that determine whether a generated sample $V(\bm{X}_{\text{out}})$ is real or fake and identify who the source speaker of $V(\bm{X}_{\text{out}})$ is. }
  
  \label{fig:slmgan}
\end{figure*}

\section{Methods}
\label{sec:methods}
\subsection{SLMGAN} 
SLMGAN builds upon StarGANv2-VC \cite{li2021starganv2}, a many-to-many voice conversion system that uses a mel-discriminator to determine whether the converted speech is real or fake, and a mel-based source classifier to identify the source speaker of the converted speech. We introduce several modifications to StarGANv2-VC: we add a SLM-based discriminator in addition to the mel-based discriminator, substitute the mel-based source classifier with the SLM-based one, remove the mapping network and domain-specific projections, and replace the ASR loss with the SLM loss. Our framework consists of four modules. Figure~\ref{fig:slmgan} provides an overview of our framework.

\noindent\textbf{Generator.} The generator $G$ transforms an input mel-spectrogram $\bm{X_{\text{src}}}$ into $G(\bm{X_{\text{src}}}, s, h_{f0})$ that reflects the speaker identity in the style vector $s$ encoded by the style encoder $S$ and the pitch represented in $h_{f0}$ provided by the convolution layers in the F0 extraction network $F$. Subsequently, a pre-trained BIGVGAN \cite{lee2022bigvgan} vocoder $V(\cdot)$ converts the generated mel-spectrogram into a waveform.

\noindent\textbf{F0 network.} The F0 extraction network $F$ is a pre-trained JDC network \cite{kum2019joint} that extracts the fundamental frequency from an input mel-spectrogram. The JDC network comprises convolutional layers followed by BLSTM units. We follow \cite{li2021starganv2} and use the convolutional output $F_{\text{conv}}(\bm{X})$ for $\bm{X} \in \mathcal{X}$ as the input features.

\noindent\textbf{Style encoder.} Given a reference mel-spectrogram $\bm{X}_{\text{ref}}$, the style encoder $S$ extracts the style vector $s = S(\bm{X}_{\text{ref}})$. We remove the domain-specific projections to generalize to the zero-shot scenario, enabling the style encoder to function as a speaker encoder. 
   
\noindent\textbf{Discriminators.} We introduce the SLM-based discriminator on top of the mel-based domain-specific discriminator and replace the source  classifier in \cite{li2021starganv2} with a SLM-based classifier. The base architecture of the discriminator and classifier is the same as that of spec-discriminators in \cite{lee2022bigvgan}. Instead of spectrograms, the input is SLM features from a pre-trained 12-layer WavLM \cite{chen2022wavlm} model fine-tuned for speaker verification tasks\footnote{Available at \mbox{\footurl{https://huggingface.co/microsoft/wavlm-base-plus-sv}}}. We downsample the input to 16 kHz and append a linear projection head of size $13 \times 768 \rightarrow 256$ as the input channels before the convolutional layers. To stabilize training, the SLM-based discriminator joins training after epoch 20.

\subsection{Training Objectives}
SLMGAN aims to learn a mapping $G:  \mathcal{X}_{y_{\text{src}}} \rightarrow \mathcal{X}_{y_{\text{trg}}}$ that converts a sample $\bm{X}\in \mathcal{X}_{y_{\text{src}}}$ from the source speaker $y_{\text{src}} \in \mathcal{Y}$ to a sample $\bm{\hat{X}} \in \mathcal{X}_{y_{\text{trg}}}$ in the target speaker $y_{\text{trg}} \in \mathcal{Y}$ without parallel data and text labels. The underlying distribution of speakers $\mathcal{Y}_{\text{train}}$ and $\mathcal{Y}_{\text{test}}$ differs, and the goal is to generalize to unseen speakers in $\mathcal{Y}_{\text{test}}$. During training, we sample a reference mel-spectrogram $\bm{X}_{\text{ref}}$ of speakers in $\mathcal{Y}_{\text{train}}$ to get its style vector $s$ via style encoder where $s = S(\bm{X}_{\text{ref}})$. Given a mel-spectrogram $\bm{X}\in \mathcal{X}_{y_{\text{src}}}$, the source domain $y_{\text{src}} \in \mathcal{Y}_{\text{train}}$ and the target domain $y_{\text{trg}} \in \mathcal{Y}_{\text{train}}$, we train our model with the following objectives:
    
\noindent\textbf{Adversarial loss.} We follow \cite{lee2022bigvgan} by using the LSGAN \cite{mao2017least} loss :
\begin{equation}
      \mathcal{L}_{adv}(G;D) =\text{ } \mathbb{E}_{\bm{X}, y_{\text{trg}}, s}\left[\parn{D(G(\bm{X}, s), y_{\text{trg}}) - 1}^2\right], \\
\end{equation}

    \begin{equation}
    \begin{aligned}    
      \mathcal{L}_{adv}(D;G) =\text{ }
      &\mathbb{E}_{\bm{X}, y_{\text{trg}}, y_{\text{src}}, s}\left[\parn{D(G(\bm{X}, s), y_{\text{trg}})}^2\right] + \\ &\mathbb{E}_{\bm{X}, y_{\text{trg}}, y_{\text{src}}, s}\left[
       \parn{D(V(\bm{X}), y_{\text{src}}) - 1}^2\right],
    \end{aligned}
      \label{eq1}
    \end{equation}
    where $D(\cdot, y)$ denotes the output of discriminator for speaker $y \in \mathcal{Y}_{\text{train}}$ and $G$ denotes the  waveform output. We retain the same cross-entropy loss in \cite{li2021starganv2} for the mel-discriminator in the first 20 epochs. 
    
    \noindent\textbf{Adversarial source classifier loss.} We employ the source classifier and adversarial source classifier loss as in \cite{li2021starganv2}: 
    \begin{equation}
      \mathcal{L}_{cls} = \mathbb{E}_{\bm{X}, y_{\text{src}}, s}\left[\textbf{CE}(C(G(\bm{X}, s)), y_{\text{src}})\right]
      \label{eq10}
    \end{equation}
    \begin{equation}
      \mathcal{L}_{advcls} = \mathbb{E}_{\bm{X}, y_{\text{trg}}, s}\left[\textbf{CE}(C(G(\bm{X}, s)), y_{\text{trg}})\right]
      \label{eq2}
    \end{equation}
    where $\text{CE}(\cdot)$ denotes the cross-entropy loss function.
    
    \noindent\textbf{Style reconstruction loss.} We use the same style reconstruction loss to make sure the converted speech has the same style vector as the reference speech:
    \begin{equation}
      \mathcal{L}_{sty} = \mathbb{E}_{\bm{X}, y_{\text{trg}}, s}\left[\norm{s - S(\tilde{G}(\bm{X}, s), y_{\text{trg}})}_1\right],
      \label{eq3}
    \end{equation}
    where $\tilde{G}(\cdot)$ denotes the mel-spectrogram of converted speech, and hence $G = V(\tilde{G})$. 
    
    \noindent\textbf{F0 consistency loss.} 
    We follow \cite{li2021starganv2} and use the F0 consistency loss to produce F0-consistent results. We normalize the absolute F0 values $F(\bm{X})$ by its temporal mean, denoted by $\hat{F}(\bm{X}) = \frac{F(\bm{X})}{\norm{F(\bm{X})}_1}$. The F0 consistency loss is
    \begin{equation}
      \mathcal{L}_{f0} = \mathbb{E}_{\bm{X}, s}\left[\norm{\hat{F}(\bm{X}) - \hat{F}(\tilde{G}(\bm{X}, s))}_1\right]
      \label{eq5}
    \end{equation}
    
    \noindent\textbf{Speech consistency loss.} 
    Instead of intermediate features from ASR models used in \cite{li2021starganv2}, we leverage the WavLM representations from layer 6 to layer 9, denoted by $h_{slm}(\cdot)$, as the linguistic features, since these layers are reported to contain the most phonetic information with little speaker information \cite{choi2021neural, pasad2021layer}. The speech consistency loss is defined as
    \begin{equation}
      \mathcal{L}_{slm} = \mathbb{E}_{\bm{X}, s}\left[\norm{h_{slm}(V(\bm{X})) - h_{slm}(G(\bm{X}, s))}_1\right]
      \label{eq6}
    \end{equation}
    
    \noindent\textbf{Norm consistency loss.} We use the norm consistency loss to preserve the speech/silence intervals of generated samples. We used the same definition of norm as in \cite{li2021starganv2}, 
    $\norm{\bm{X}_{\cdot, t}} = \sum\limits_{n = 1}^N |\bm{X}_{n, t}|$, where $t \in \{1, \ldots, T\}$ is the frame index. The norm consistency loss is 
    \begin{equation}
      \mathcal{L}_{norm} = \mathbb{E}_{\bm{X}, s} \left[\frac{1}{T}\sum\limits_{t =  1}^T\abs{\norm{\bm{X}_{\cdot, t}} - \norm{\tilde{G}(\bm{X}, s)_{\cdot, t}}}\right]
      \label{eq7}
    \end{equation}
    
    \noindent\textbf{Cycle consistency loss.} Lastly, we employ the cycle consistency loss as in \cite{li2021starganv2} to preserve all other features of the input
    \begin{equation}
      \mathcal{L}_{cyc} = \mathbb{E}_{\bm{X}, y_{\text{src}}, y_{\text{trg}}, s} \left[\norm{\bm{X} - \tilde{G}(\tilde{G}(\bm{X}, s), \tilde{s}))}_1\right]
      \label{eq8}
    \end{equation}
    where $\tilde{s} = S(\bm{X})$ is the style vector of the source speaker $y_{\text{src}}$.
    
    \noindent\textbf{Full objective.} Our full generator objective functions can be summarized as follows:
    \begin{equation}
    \begin{aligned}
        \min_{G, S} &\text{  }
       \mathcal{L}_{adv}(G;D) + \lambda_{advcls}\mathcal{L}_{advcls}  + \\
       &\lambda_{sty}  \mathcal{L}_{sty} + 
        \lambda_{f0}  \mathcal{L}_{f0} + 
      \lambda_{slm}  \mathcal{L}_{slm} \\
      &+\lambda_{norm}  \mathcal{L}_{norm} +
      \lambda_{cyc}  \mathcal{L}_{cyc}
    \end{aligned}
      \label{eq9}
    \end{equation}
    where $\lambda_{advcls}, \lambda_{sty}, \lambda_{f0}, \lambda_{slm}, \lambda_{norm}$ and $ \lambda_{cyc}$ are hyperparameters for each term. 
    
    Our full discriminators objective is given by: 
    \begin{equation}
        \min_{C, D} \mathcal{L}_{adv}(D;G) + \lambda_{cls}\mathcal{L}_{cls} 
      \label{eq10}
    \end{equation}
    where $\lambda_{cls}$ is the hyperparameter for source classifier loss $\mathcal{L}_{cls}$.

\section{Experiments}
\label{sec:experiments}

\subsection{Datasets} We evaluated our models using the VCTK \cite{yamagishi2019cstr} corpus. The VCTK dataset comprises 109 native English speakers with various accents, each reading around 400 sentences. We followed the procedure in \cite{wang2021vqmivc, li2023styletts}, selecting 89 speakers randomly for training and using the remaining 20 speakers as unseen speakers for testing. We  divided the samples of the 89 speakers into training and validation sets with a 90\%/10\% split. The samples were downsampled to 22.5 kHz to match the sampling rate of pre-trained BIGVGAN.

\subsection{Training Details} We trained our model for 90 epochs, with a batch size of 28 two-second long audio segments. We used the same hyperparameters as in \cite{li2021starganv2}, with $\lambda_{cls} = 0.1, \lambda_{advcls} = 0.5, \lambda_{sty} = 1, \lambda_{f0} = 5, \lambda_{slm} = 1, \lambda_{norm} = 1$ and $ \lambda_{cyc} = 1$. We trained our model using the AdamW optimizer \cite{loshchilov2018fixing} with $\beta_1 = 0, \beta_2 = 0.99$, weight decay $\lambda = 10^{-4}$, and learning rate $\gamma = 10^{-4}$. The SLM-based discriminator joined training after epoch 20 to accelerate and stabilize training. We also used bCR regularization \cite{zhao2021improved} after epoch 20, as in \cite{li2021starganv2}. The source classifier training started after epoch 35.

\begin{table}[!t]
\vspace{-0.2cm} 

	\centering
	\caption{MOS with 95\% confidence intervals between different models. All results are statistically significant relative to SLMGAN.  \\ } 
    \begin{tabular}{c|c|c}
    \hline
    Method & MOS-N & MOS-S \\
    \hline
    Ground Truth      & 4.24 ($\pm$ 0.06)  & 4.28 ($\pm$ 0.08) \\
    \hline
    SLMGAN   & \textbf{3.88} ($\pm$ \textbf{0.12})  &    {3.29} ($\pm$ {0.14}) \\
    StyleTTS-VC   & {3.62} ($\pm$ {0.13})  &    \textbf{3.55} ($\pm$ \textbf{0.13}) \\
    YourTTS & 3.54 ($\pm$ 0.13) & 3.49 ($\pm$ 0.14) \\
    VQMIVC  & 2.58 ($\pm$ 0.10) & 2.43 ($\pm$ 0.09) \\
    AGAIN-VC & 2.02 ($\pm$ 0.07)  & 2.07 ($\pm$ 0.10) \\
    \hline
    \end{tabular}
    \label{tab:t1}
\vspace{-0.2cm} 

\end{table}

\begin{table}[!th]
\vspace{-0.2cm} 
	\centering
	\caption{Objective evaluation results of test speaker classification accuracy (ACC), phoneme error rate (PER), and real time factor (RTF) between different models. The RTF was calculated under a single NVIDIA GeForce RTX 3090 Ti GPU.\\ } 
    \begin{tabular}{c|c|c|c}
    \hline
    Method & ACC $\uparrow$ & PER $\downarrow$ & RTF $\downarrow$  \\
    \hline
    Ground Truth      & 100\% & 3.1\% & -- \\
    \hline
    SLMGAN   & {68.9\%}  & {7.8\%} & \textbf{0.0076} \\
    StyleTTS-VC   & \textbf{90.8\%}  & {6.52\%} & 0.0128\\
    YourTTS & 51.3 \% & \textbf{5.49\%}   & 0.0375 \\
    VQMIVC & 31.4\% & 26.5\% & {0.0117}\\
    AGAIN-VC & 70.5\% & 24.9\% & 0.0138\\

    \hline
    \end{tabular}
    \label{tab:t2}
    \vspace{-0.3cm} 

\end{table}

\subsection{Evaluations}
\label{section3.3}
We conducted subjective evaluations using two metrics: the mean opinion score of naturalness (MOS-N), which measures the naturalness of converted speech, and the mean opinion score of similarity (MOS-S), which assesses the similarity between converted and reference speech. We recruited native English speakers in the U.S. to participate in our evaluations using an online survey through Amazon Mechanical Turk. We compared our model with two recent baseline models, AGAIN-VC \cite{chen2021again} and VQMIVC \cite{wang2021vqmivc}, and two state-of-the-art models, YourTTS\cite{casanova2022yourtts} and StyleTTS-VC \cite{li2023styletts}, for zero-shot voice conversion. All baseline models were trained with the official implementation 
using the same train and test speaker split. Since the official recipe for YourTTS was in 16 kHz, we downsampled outputs from all other models to 16 kHz in our evaluations.

In each experiment, we randomly selected 40 sets of samples. For each set, we demanded at least four different speakers in VCTK to read the same sentence, with one serving as the ground truth and the remaining three used as the source input for our model and the two baseline models. Since there are not enough pairs to support all six models, we made two surveys, one with our model, ground truth, StyleTTS-VC, and YourTTS, and another with our model, ground truth, VQMIVC, and AGAIN-VC. This ensures that different samples have varying speaking speeds and tones, preventing raters from identifying the ground truth. When evaluating each set, we randomly permuted the order of the models without revealing the model labels, allowing the raters to compare the subtle differences among models, similar to multiple stimuli with hidden reference and anchor (MUSHRA)  \cite{li2022styletts}. We used the subjective rating of the ground truth as an attention check: all ratings from a subject were discarded from our analyses if the MOS of the ground truth was not ranked the highest among all models. Each set was rated by 10 raters after disqualifying raters who failed the attention check.

We also performed objective evaluations. We used a pre-trained speaker classification model to measure speaker classification accuracy (ACC) as a metric for speaker similarity and an ASR model to measure phoneme error rate (PER) for speech intelligibly. The speaker classification model is a ResNet-18 network that takes 80-dimensional log filter banks as the mel-spectrogram input to predict the speaker label \cite{li2023styletts}. The model was trained on the test speakers, and we report the classification accuracy (ACC) of the trained models on samples generated with different models. We transcribed speech waveforms to text using an ASR model from ESPNet \cite{watanabe2018espnet} and converted the text to phoneme sequences to calculate PER.

\subsection{Ablation Study}
\label{section3.4}
To show that our approach of using SLM-based discriminators is effective, we conducted an ablation study with subjective and objective evaluations described in section \ref{section3.3}. We replaced $\mathcal{L}_{slm}$ with $\mathcal{L}_{asr}$ as in \cite{li2021starganv2} ($\mathcal{L}_{slm} \rightarrow \mathcal{L}_{asr}$ in Table \ref{tab:3}) and replaced the SLM-based discriminators with mel-based ones as in \cite{li2021starganv2} (SLM $\rightarrow$ Mel in Table \ref{tab:3}), to demonstrate the effectiveness of our newly designed speech consistency loss and SLM-based discriminators.

\begin{table}[!t]
    \vspace{-0.2cm} 

	\centering
	\caption{Comparison of MOS with 95\% confidence intervals (CI), test accuracy (ACC) and PER in the ablation study. \\ } 
 \begin{adjustbox}{width=\columnwidth,center}
    \begin{tabular}{c|c|c|c|c}
    \hline
    Method & MOS-N & MOS-S & ACC & PER \\
    \hline
    Baseline      &  \textbf{3.96} ($\pm$ \textbf{0.09})  &  \textbf{3.40} ($\pm$ \textbf{0.11}) & \textbf{68.9} \%  &7.8\% \\
    SLM $\rightarrow$ Mel   & 2.68 ($\pm$ 0.11)  & 2.82 ($\pm$ 0.10) & \textbf{68.9} \%  & \textbf{6.2} \%\\
     $\mathcal{L}_{slm} \rightarrow \mathcal{L}_{asr}$ & 2.64 ($\pm$ 0.10)  & 2.67 ($\pm$ 0.10) &64.4\%  &10.7\%\\
    \hline
    \end{tabular}
    \end{adjustbox}
    \vspace{-0.2cm} 

    \label{tab:3}
    
\end{table}
\section{Results}
\begin{figure}
    \centering
    \includegraphics[width=\columnwidth]{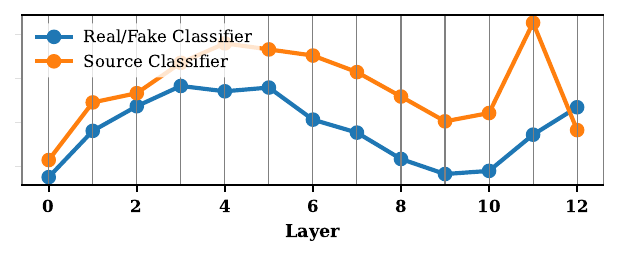}
    \vspace{-0.7cm} 
    \caption{Normalized weight magnitude (importance) of the projection head for SLM features in different layers.}
    \label{fig:my_label}
    \vspace{-0.4cm} 

\end{figure}
\label{sec:results}
Table \ref{tab:t1} presents the comparison results between different models and the ground truth. Our model significantly outperforms other models in terms of naturalness in the subjective evaluation experiments, establishing a new state-of-the-art. Although StyleTTS-VC and YourTTS achieved higher similarity scores, it is important to note that our model requires less information than these two TTS-based models, as we do not need text labels during training. Despite this, our model still demonstrates performance comparable to these two models and surpasses VQMIVC and AGAIN-VC in terms of similarity. Our model also achieves comparable PER and ACC in Table \ref{tab:t2} to YourTTS, even without text labels during training. Furthermore, our model is significantly faster than all other models, being nearly $2\times$ faster than StyleTTS-VC and $5\times$ faster than YourTTS, making it an attractive choice for real-time applications.

Table \ref{tab:3} shows that SLM-based discriminators and SLM-based speech consistency loss significantly improve both similarity and naturalness, as well as speech clarity, as indicated by the PER. It is also worth noting that removing these two components effectively reverts the model back to the original design of StarGANv2-VC. However, the performance declines dramatically compared to the results reported in \cite{li2021starganv2} when the number of speakers is large. This suggests that mel-based discriminators and ASR loss cannot scale well with a larger number of speakers as they were not pre-trained with a larger number of  speakers. These findings highlight an alternative application of SLM representations, utilizing them as training objectives instead of latent features from which speech is decoded.

Figure \ref{fig:my_label} displays the weight importance across layers of the projection head in the SLM-based discriminators. The importance peaks around earlier layers, indicating that discriminators rely more on acoustic and phonetic features, while semantic information from later layers is less critical for the discriminative tasks. Unlike \cite{li2023styletts2}, because the WavLM model is fine-tuned for speaker verification tasks, the last two layers, which are directly optimized, also have the largest weight importance. This insight can guide future research in selecting optimal SLMs based on their pretext and fine-tuned tasks.

\section{Discussions}
\label{sec:discussions}
In this work, we introduced SLMGAN, a zero-shot voice conversion (VC) model with SLM-based discriminators and speech consistency loss. Our model achieves state-of-the-art naturalness and competitive similarity without text labels and outperforms baseline models in inference speed. The ablation study highlights the effectiveness of SLM-based components and suggests their potential as training objectives for voice conversion. Layer-wise weight analysis informs future optimal SLM choices. Future direction includes boosting speaker similarity and applicability to other cross-domain voice transfer like emotion and accent conversion. In addition, the training setup with big vocoder models such as BIGVGAN can be burdensome, so future works could also benefit from developing high-quality lightweight vocoders for training.

\section{Acknowledgments}
\label{sec:acknowledgments}
This work was funded by the national institute of health (NIHNIDCD) and a grant from Marie-Josee and Henry R. Kravis. 

\newpage
\bibliographystyle{IEEEtran}
\bibliography{refs23}
%
%
%
%
%
%
%
%
%

\end{sloppy}
\end{document}